\documentclass[aps,pre,url,twocolumn 
,superscriptaddress]{revtex4-2}

\usepackage{latexsym}
\usepackage{amssymb,amsmath,amsbsy}
\usepackage{graphicx,color}
\usepackage{bm}
\usepackage{longtable}
\usepackage{epsf}
\usepackage[utf8]{inputenc}
\usepackage{enumerate}
\usepackage[english]{babel}               

\newcommand{\Fcal}{\mathcal{F}}

\newcommand{\Scal}{\mathcal{S}}

\newcommand{\D}{\mathcal{D}}

\newcommand{\Z}{\mathcal{Z}}
\newcommand{\s}{\mathcal{S}}
\newcommand{\XX}{\mathbb{X}}
\newcommand{\PP}{\mathbb{P}}
\newcommand{\TT}{\mathbb{T}}
\newcommand{\FF}{\mathbb{F}}

\newcommand{\eRM}{\mathrm{e}}
\newcommand{\dd}{\mathrm{d}}

\newcommand{\vv}{{\bm v}}

\newcommand{\G}{\Gamma}

\newcommand{\xx}{{\bm x}}
\newcommand{\yy}{{\bm y}}
\newcommand{\kk}{{\bm k}}
\newcommand{\pp}{{\bm p}}
\newcommand{\qq}{{\bm q}}

\newcommand{\rr}{{\bm r}}

\newcommand{\eps}{\varepsilon}

\allowdisplaybreaks

\usepackage{comment}

\begin{document}


\title {Universality in Incompressible Active Fluid: \\ Effect of Non-local Shear Stress}
\author{V.~\v{S}kult\'ety}
\email{viktor.skultety@ed.ac.uk}
\affiliation{SUPA, School of Physics and Astronomy, The University of Edinburgh, Peter Guthrie Tait Road, Edinburgh EH9 3FD, United Kingdom
}
\author{\v{S}. Birn\v{s}teinov\'a} 
\affiliation{Faculty of Sciences, P.~J.~\v{S}af\'arik University, 04154 Ko\v{s}ice, Slovakia}
\author{T.~Lu\v{c}ivjansk\'y}
\affiliation{Faculty of Sciences, P.~J.~\v{S}af\'arik University, 04154 Ko\v{s}ice, Slovakia}
\author{J. Honkonen} 
\affiliation{Department of Military Technology, National Defence University, P.O.Box 7,00861, Helsinki, Finland}
\date{\today}
\begin{abstract}
	

	Phase transitions in active fluids attracted significant attention within the last decades. Recent results show [L. Chen \emph{et al.}, New J. Phys. 17, 042002 (2015)] that an order-disorder phase transition in incompressible active fluids belongs to a new universality class. In this work, we further investigate this type of phase transition and focus on the effect of long-range interactions. This is achieved by introducing a non-local shear stress  into the hydrodynamic description, which leads to superdiffusion of the velocity field, and can be viewed as a result of the active particles performing L\'evy walks. The universal properties in the critical region are derived by performing a perturbative renormalization group analysis of the corresponding response functional within the one-loop approximation. We show that the effect of non-local shear stress decreases the upper critical dimension of the model, and can lead to the irrelevance of the active fluid self-advection with the resulting model belonging to an unusual 'long-range Model A' universality class not reported before. Moreover, when the degree of non-locality is sufficiently high all non-linearities become irrelevant and the mean-field description is valid in any spatial dimension.	
\end{abstract}

\maketitle

\section{Introduction \label{sec:intro}}

In recent years, many biological systems received significant interest because of their rich non-equilibrium statistical physics properties \cite{MJRLPRS13,Cates2012}. 
One of the current challenges is to describe the behavior of a large number of active particles, such as fish, birds, bacteria, etc. The main distinction of active matter systems from   their passive counterparts is the ability of their particles to extract the energy from the smallest spatial scales, e.g.~consume food resources in an ambient environment, and   utilize it for the self-propulsion. Such non-equilibrium systems can exhibit intriguing macroscopic collective behavior, which might be observed in almost every realm in nature~\cite{TTR05,Voituriez2005,Rappel1999,Sokolov2012,Wensink2012,DHBG13,Gachelin2014}.  Whereas biologists tend to build detailed representations of a particular case, the ubiquity of the phenomenon suggests underlying universal features and thus gives weight to the bottom-up modelling approach 
usually favored by the physics community.

The theoretical study of active systems was pioneered three decades ago by the work of Vicsek et al.~\cite{VCB95}, on the flocking behavior. Using numerical simulations, a continuous order-disorder phase transition was claimed to be found in the two-dimensional system of self-propelled particles with alignment interactions.
It was later shown, that the precise origin of the phase transition is very sensitive to details of the simulation: For example the system size or different protocols for velocity update might change the type of phase transition from second to first order \cite{Gregoire:2004:PRL,nagy07,baglietto08,Chate2008,Vicsek2012}. Nevertheless, 
 the Vicsek model brought about a new 
and fruitful
  direction in the active matter research. Since then, various theoretical and experimental active systems were studied, from the polar ordered active matter \cite{TTR05} and systems with motility inducted phase separation \cite{CT15} to molecular motors \cite{Sanchez12}, light activated colloids \cite{Palacci2013}, birds and fish  swarms~ \cite{PH97,Cavagna2018,Cavagna2019,Cavagna2019a}. Marchetti et al.~\cite{MJRLPRS13} recently classified active systems into two classes depending on the  interaction between particles - dry and wet active matter. In the first case, particles interact mainly with the surrounding medium via effective friction force, such as Viscek-like particles \cite{Vicsek2012,Chate2020}, while the latter is dominated by the long-range hydrodynamic interactions between particles \cite{SS13,Saintillan2018,Stenhammar2017,Skultety2020}. 

Historically, the first studied continuous model was a dry active matter system introduced by Toner and Tu \cite{TT95}.  They introduced the coarse-grained hydrodynamic equation for the active medium  using phenomenological and symmetry considerations. The hydrodynamics is described by a Navier-Stokes-like equation with violated Galilean invariance, which can be interpreted as a direct consequence of the lacking momentum conservation. The order-disorder phase transition was extensively investigated by the means of renormalization group (RG) method \cite{TT95,TT98,TTU98,TTR05}. However, it was later realized that the ordered state is actually unstable to any density perturbations along the mean velocity field \cite{bertin2006,bertin2009}. This led Chen \emph{et al.} \cite{CTL15} to analyze the incompressible version of Toner-Tu model. They showed that its order-disorder transition belongs to a new universality class. One should note that the fluid's incompressibility can not only be simulated \cite{Wensink2012,Ramaswamy2015}, but it has also been reported to arise in various experimental situations, such as dense systems with strong repulsive short-range interactions \cite{Wensink2012} or systems with long-range interactions \cite{Pearce2014,Steimel2016}. 
Moreover, experimental findings show very small speed fluctuations in starling flocks \cite{Cavagna2010,Attanasi2014}. At last, the incompressible Tonner-Tu model is not only limited to description of flocks and swarms, but its edge instability has also been considered to describe tissue regeneration in plants \cite{Nesbitt2017}.

Despite the appearance of the long-range interactions in some active systems, the original incompressible Toner-Tu model does not assume any  non-local interaction apart from the incompressibility condition. This motivates the present work, in which we propose a modification of active fluid equations, by inclusion of non-local shear stress into the hydrodynamic equations of motion. As we will show, the non-local shear stress results in the superdiffusion of the velocity and the vorticity field. The origin of this behaviour can be linked to the active particles performing L\'evy walks \cite{CL19a,CL19b},  or sensing perturbations of in the velocity over longer distances \cite{Pearce2014}. In fact, superdifussive  L\'evy walks have been reported in several recent experiments~\cite{Cavagna2013,Murakami2015,Ariel2015,KPV19}. Furthermore, our minimal extension of the original   model is the most convenient for practical calculations, since it leads to the appearance of the fractional Laplace operator, whose renormalization is well  understood in field theories ~\cite{HN89,Janssen2008,Honkonen2018}.

Recently, there has also been an increased interest in dense active matter systems
that might display glassy-like behavior~\cite{henkes11,bialke12,bechinger16,sussman18,Janssen_2019}. In particular, closely related aging phenomenon
is gaining more and more attention. 
Necessary condition for aging is a presence of broken time-invariance, e.g. by a preparing system 
with special initial conditions~\cite{henkelbook,Janssen_2019}. However, in this
work our aim is to study steady state and therefore we choose such formulation of model in which
 time-invariance is preserved. 

This work is divided into three sections and two appendices. In Sec.~\ref{sec:class}  we introduce the incompressible Dry active matter model analyzed by Chen et al. \cite{CTL15}. The model is then recast into the De Dominicis-Janssen response   functional formalism and the perturbation theory is described. The analysis of the ultraviolet (UV) divergences is performed and the renormalization process is carried out   within the   leading  one-loop approximation by the means of dimensional regularization and subsequent $ \varepsilon $-expansion, consistent with the former calculations~\cite{CTL15}.   In Sec.~\ref{sec:nonlocal} we introduce the non-local shear stress into the hydrodynamic description of the model and briefly discuss its interpretation, and implication on the energy dissipation. We then   discuss the non-local modification of the response functional formalism and carry out the renormalization procedure   once more, however, in different way by the combined analytic-dimensional renormalization scheme. We finalize our field-theoretic    analysis with the calculation of fixed points,    their stability and corresponding critical exponents. Final remarks are reserved for concluding Sec.~\ref{sec:conclusion}. Details of the actual calculations and explicit results of    RG functions can be found in the appendices~\ref{app:FD} and \ref{app:RC}, respectively.

{ \section{Classical incompressible active fluid} \label{sec:class}}

{\subsection{Mesoscopic description \label{subsec:micForm}}}
Let us stress once again that in this work we employ field-theoretic renormalization  group (RG) method with dimensional regularization of Feynman diagrams.  This presupposes use of space dimension $d$ as a complex variable \cite{Zinn,collins_1984}. Hence, whenever necessary we retain and write $d$-dependence explicitly.  For future convenience we also abbreviate spatio-temporal coordinates in the  following way $x \equiv (t,\xx)=(t,x_1,\ldots,x_d)$.

The coarse grained description of the incompressible polar dry active fluid is based on general hydrodynamic considerations \cite{Martin72,Lubensky95,Wensink2012,DHBG13}. These are based on a proper identification of slow variables and symmetries present in the model.  The starting point is the generalized stochastic equation of motion, which takes the following form
\begin{align}
  \partial_{t} v_{i} + v_{j}\partial_{j} v_{i} &=
  \partial_{j} \mathcal{E}_{ji}    - \partial_{i} p + \mathcal{F}_{i} + f_{i}, 
  \label{eq:eom1}
  \\
  \partial_{i} v_{i} &= 0,
  \label{eq:eom2}
\end{align}
where $ \partial_{t} \equiv \partial/\partial t$ is time derivative, $\partial_{j} \equiv \partial/\partial x_{j} $ is spatial derivative, $  v_{i} \equiv v_{i}(x) $ is $i$-th component of the velocity field $\vv=(v_1,\ldots,v_d)$. In contrast to the original Toner-Tu model~\cite{TT95,TT98,Toner2012},  we assume that the system is incompressible, which leads to Eq.~\eqref{eq:eom2}. The parameter $\lambda_0$ not present   in the Navier-Stokes equation is allowed in the convective term in Eq.~\eqref{eq:eom1}, because the model is not Galilean invariant and is valid only in a special coordinate system in which the background environment is fixed. Hence, in general we have $ \lambda_{0} \neq 1 $.
  
 Further, $\mathcal{E}_{ij} $ stands for the strain rate tensor, and $ p \equiv p(x) $ is the pressure field which acts as a Lagrange multiplier that enforces the incompressibility condition. Since we are interested in the dry active fluid the total momentum is not conserved. The simplest choice for the force $ \Fcal_i $ in accordance
  with the symmetries is to introduce it in  the following way
\begin{equation}
  \Fcal_{i} = 
  - \frac{\delta}{\delta v_{i}} U[\vv], 
  \quad U[\vv] =
  \frac{1}{2} \tau_{0} |\vv|^{2} + \frac{1}{4!} g_{10} |\vv|^{4},
  \label{eq:def_force}
\end{equation}
 where $U[\vv]$ is akin to Landau potential in critical dynamics, parameters $ \tau_{0} $ and $ g_{0} $ are the mesoscopic deviation from the criticality and the coupling constant, respectively. 

 The above force $ \mathcal{F}_{i} $ might be interpreted as a friction force of the environment acting on the active particles. On the  other hand, the random force $ f_{i} $ is responsible for the continuous energy supply from the microscopic spatial scales and mimics  the inherent  stochasticity of the process. As usual, $f_i$ is assumed to obey Gaussian statistics with zero mean and two point correlator
\begin{equation}
  \langle f_{i}(x) f_{j}(x') \rangle = \delta(t-t') \int \dd^{d} k
  \ D_{ij}^{v}(\kk) \eRM^{i\kk\cdot\rr}, 
\end{equation}
where $ \rr = \xx - \xx' $ and the kernel function $D_{ij}^v$ takes the following form
\begin{equation}  
  D_{ij}^{v}(\kk)= \nu_{0} \PP_{ij}(\kk).
\end{equation}
Hereinafter, $ \PP_{ij}(\kk) = \delta_{ij} - k_{i}k_{j}/k^{2} $ denotes the transversal projection operator in the momentum representation introduced because any longitudinal component of the random noise $ f_{i} $ can be eliminated by the redefinition of the pressure field.

In accordance with previous works~\cite{CTL15,DHBG13}  we postulate components of the modified strain rate tensor to be
\begin{equation}
  \mathcal{E}_{ij} = \nu_{0} \varepsilon_{ij} + S_{0} Q_{ij},
\end{equation}
where $ \varepsilon_{ij} = ( \partial_{i} v_{j} + \partial_{j} v_{i}) $ is the classical strain rate tensor, $ Q_{ij} = v_{i} v_{j} - \delta_{ij} |\vv|^{2} / d $  is the active nematic stress tensor \cite{SR02,Lubensky95} and $ \nu_{0}, S_{0} $ are the microscopic viscosity and the microscopic amplitude. The ensuing incompressible Toner-Tu model takes the following form~\cite{CTL15}
\begin{align}
  \label{SR:eq_ANS}
  \partial_{t} v_{i} + \lambda_{0} v_{j}\partial_{j} v_{i} = &\ 
  \nu_{0} \partial^{2} v_{i} -  
   \left( \tau_{0} + g_{10}|\vv|^{2}/3! \right) v_{j}
  \nonumber \\
  &\ - \partial_{i} \tilde{p}  + f_{i},  \\
  \partial_{i} v_{i} =&\ 0, 
\end{align}
where $ \lambda_{0} = 1 - S_{0}  $ and $ \tilde{p} = p + (S_{0}/d) |\vv|^{2} $ is the modified pressure term, which does not affect the  universal properties of the system due to the incompressibility condition.

In contrast to incompressible fluid the energy balance equation for the active fluid contains two additional terms
\begin{equation}
  \partial_{t} E = - 2\nu \langle\langle \omega^{2} \rangle\rangle - 2 \tau_{0} E - \frac{1}{3!} 
  g_{10} \langle\langle v^{4} \rangle\rangle + \langle\langle f_{i}v_{i} \rangle\rangle, 
  \label{1:eq.TTc}
\end{equation}
where $ E = \langle\langle v^{2}/2 \rangle\rangle $ is the total kinetic energy, $ \omega_{i} = \varepsilon_{ijk} \partial_{j} v_{k} $ is the vorticity and $ \langle\langle \dots \rangle\rangle $ stands for an integration over spatial variable $\xx$, i.e.
\begin{equation}
   \langle\langle \ldots \rangle\rangle \equiv \int \dd^d x \ldots .
\end{equation}     
    In the disordered phase the parameter $\tau_0$ takes on positive values, and only the random force is responsible for the energy input. In the case of stationary flow $ \partial_{t} E  = 0 $, the random  force has to compensate not only for the dissipation of the energy caused by the viscous forces, but      by the "friction forces" as well. 

\subsection{Field-theoretic formulation \label{subsec:FT}}
Following standard procedures \cite{Vasilev04,Tauber}, we derive De Dominicis-Janssen response functional for the incompressible dry active fluid 
\begin{align}
  \Scal^{\text{SR}}[\vv',\vv] =&\ v_{i}'
   \Big\{ \partial_{t} + \nu_{0}(-\partial^{2} + \tau_{0}) \Big\} v_{i}
   \nonumber\\
   & \ +
  \nu_{0} v_{i}' 
  \Big( \lambda_{0}  v_{j} \partial_{j} + g_{10} |\vv|^{2} / 3!
  \Big) v_{i} \nonumber \\
   &\ - v_{i}' D_{ij}^{\vv} v_{j}'/2,
  \label{1:eq.SSR} 
\end{align}
where we have rescaled the parameters as
\begin{equation} 
  \tau_{0} \rightarrow \nu_{0} \tau_{0}, \ g_{10} \rightarrow \nu_{0}
 g_{10}, \ \lambda_{0} \rightarrow \nu_{0} \lambda_{0} ,
\end{equation} 
  due to the dimensional reasons. Note, that the modified pressure term has disappeared from Eq.~\eqref{1:eq.SSR} due to the transversality of the response field $ \vv' $. We have also used a condensed notation, in which integrals over the  spatial variable ${\xx}$ and the time variable $t$, as well as summation over repeated indices, are implicitly assumed. For instance, the second term on the right hand side of Eq.~\eqref{1:eq.SSR} corresponds to the following expression
\begin{equation}
  \label{eq:quadlocal2}
  v'_i \partial_t v_i = \int \dd t\int \dd^d{x}  
  \sum_{i=1}^d v_i'(x)\partial_t v_i(x).
\end{equation}
To finalize theoretical setup we further assume that fields $v_i$ and $v'_i$ vanish in the limits $t\rightarrow -\infty$
and $r=|\rr|\rightarrow \infty$ for any time instant $t=$ const. This corresponds to a standard formulation of initial conditions
in critical dynamics~\cite{Vasilev04,Tauber} and facilitates  use of field-theoretic approach.
 In principle, 
 it would be possible to generalize the model to account for broken time invariance~\cite{henkelbook}. However, this would
 lead to much more involved technical difficulties and such problem is left for future.  

\begin{figure}[t!]
	\centering
	\includegraphics[width=8cm]{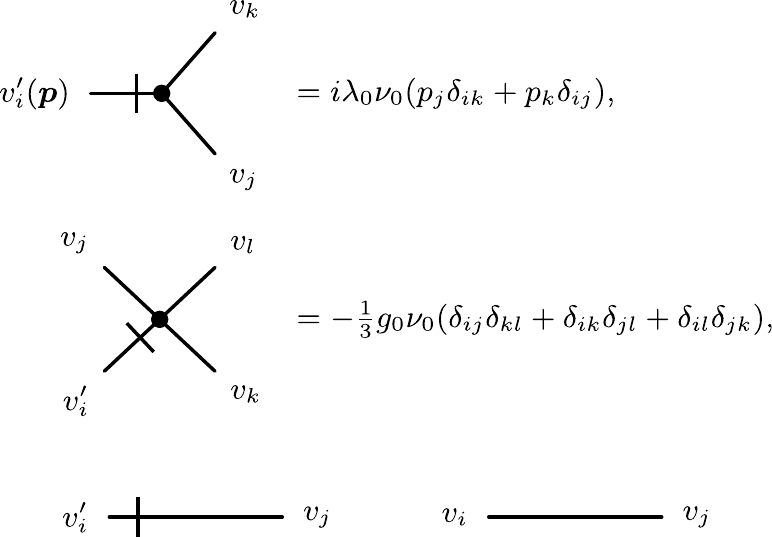} 
	\caption{Feynman rules for the incompressible active fluid model.} 
	\label{Fig1:FDR}
\end{figure}

The field-theoretic formulation implies, that all the correlation and response functions can be calculated from 
the generating functional
\begin{align} 
  \Z[A] & = \int \D \varphi \ \exp\{ -\s[\varphi] + \varphi {A} \}, \label{1:eq.Z}
  \\
  \varphi & \equiv \{ \vv',\vv \}, \quad A \equiv \{ \bm{A}^{v'}, \bm{A}^{v} \},  
\end{align}
by taking appropriate variational derivatives with respect to the corresponding source field ${A} $. For example, the linear response function is obtained as follows
\begin{align}
  \langle v_{i}(x) v_{j}'(x') \rangle = 
  \frac{\delta^{2} \Z[ A ]}{\delta A_{i}^{v'} 
  (x) \delta A_{j}^{v'}(x')}.
\end{align}
The last term in exponential of Eq.~\eqref{1:eq.Z} should be interpreted as
 scalar product between corresponding terms, i.e.
\begin{equation*}
  \varphi {A} \equiv \vv \cdot  {\bm A}^{v} + \vv' \cdot {\bm A}^{v'}.
\end{equation*}
In general, interacting field-theoretic models such as \eqref{1:eq.Z} are not exactly solvable and one may 
treat them within some perturbation scheme. Here, we utilize perturbative renormalization-group approach. For its effective use it is  advantageous to work with the effective potential $ \Gamma $, which is defined by means  of a functional Legendre transformation \cite{Vasilev04,Zinn} of   the generating functional~\eqref{1:eq.Z} 
\begin{align}
  \Gamma[\Phi] = \ln \Z[ {A} 	] - {A} \Phi, \quad \Phi(x) = 
  \frac{\delta \ln \Z[{A}]}{\delta {A}(x)}.
\end{align}
It can be shown~\cite{Vasilev04,Zinn}  that action functional and effective potential are related by the following  relation
\begin{align}
  \G[\Phi] = -\s[\Phi] + (\text{loop corrections}).
\end{align}
Effective potential $\Gamma$ also serves as a generating functional for vertex
 (one-particle irreducible) functions. These can be obtained by taking sufficiently
 many variational derivatives of $\Gamma$, i.e.
\begin{align} 
   \G^{ (N_\Phi, N_{\Phi'}) }(\{ x_i \},\{ x'_j\})  =& \ -
   \frac{\delta^{N_\Phi+N_{\Phi'}} \s[\Phi]}{\delta \Phi(x_1)  \cdots \delta \Phi'(x_1')\cdots}
   \nonumber\\ 
   & \ + \text{loop diagrams}.  
   \label{1:eq.G}
\end{align}
Here, $N_\phi$ is a number of all fields appearing in 1PI function, whereas $N_{\Phi'}$ is a number
of all response fields. 
The remaining term in Eq.~\eqref{1:eq.G}
comprises of all one-irreducible Feynman diagrams that can be constructed  using corresponding Feynman rules. As usual in field-theoretic models \cite{Tauber}, the propagators are read off from the free (quadratic) part, and the vertex factors from the interaction part of the response functional \eqref{1:eq.SSR}. The exact form of propagators and interaction vertices is the following
\begin{align}
  \langle v_{i} v_{j}' \rangle_{0} (\kk,\omega) &= 
  \frac{\mathbb{P}_{ij}(\kk)}{-i\omega + \nu_{0} (k^{2} + \tau_{0})}, \\
  \langle v_{i} v_{j} \rangle_{0} (\kk,\omega) &= \frac{\nu_{0}\mathbb{P}_{ij}
  (\kk)}    {\omega^{2} + \nu_{0}^{2}(k^{2} + \tau_{0})^{2}}, \\
  V_{v_{i}'(\pp)v_{j}v_{k}}^{\text{III}} &= i \lambda_{0} \nu_{0} \mathbb{T}_{ijk}(\pp), \\ 
  V_{v_{i}'v_{j}v_{k}v_{l}}^{\text{IV}} &=-g_{10} \nu_{0} \mathbb{F}_{ijkl},
\end{align}
where two tensor quantities 
\begin{align}
\mathbb{T}_{ijk}(\pp) &= p_{j}\delta_{ik} + p_{k} \delta_{ij}, \label{1:eq.T1} \\ 
\mathbb{F}_{ijkl} &= \frac{1}{3} (\delta_{ij}\delta_{kl} + \delta_{ik}\delta_{jl} + \delta_{il}\delta_{jk}), \label{1:eq.T2}
\end{align}
were introduced.
 A graphical representation of perturbation elements is displayed in Fig.~\ref{Fig1:FDR}. From a formal 
  point of
  view, model \eqref{1:eq.SSR} represents an 
  appealing combination of $ \phi^{4} $ and $ \partial \phi^{3} $ theory.

\subsection{Renormalization \label{subsec:Renormalization}}
\begin{table}
	\def\arraystretch{1.5}
	\centering
	\begin{tabular}{| c | c | c | c | c | c | c | c |}
		\hline
		$Q$ & ${\bm v}$ & ${\bm v}'$ & $\sqrt{\tau},\mu,\Lambda$ & $\nu_0,\nu $ & $g_{10},g_{20}=\lambda_{0}^{2}$ & $\lambda,g_{1},g_{2}$
		\\  \hline
		$d_Q^k$ & $\frac{d}{2}-1$ & $\frac{d}{2}+1$ & $1$ & $-2$ & $\varepsilon$ & $0$
		\\  \hline
		$d^\omega_Q$ & $0$ & $0$ & $0$ & $1$ & $0$ & $0$
		\\  \hline
		$d_Q$ & $\frac{d}{2}-1$ & $\frac{d}{2}+1$ & $1$ & $0$ & $\varepsilon$ & $0$
		\\ \hline    
	\end{tabular}
	\caption{Canonical dimensions of the bare fields and bare parameters for the active fluid model. The parameter $\varepsilon=4-d$.}
	\label{tab:canon_v}
\end{table}

The perturbative RG analysis is based on the analysis of the canonical dimensions of the model \cite{Vasilev04,Tauber,Zinn}.
 As a rule, dynamical models necessarily exhibit
two-scale dependence, one temporal scale and another spatial scale. 
Due to this observation total canonical dimension $ d_{Q} $ for a quantity $ Q $
is introduced
\begin{align}
  d_{Q} = d_{Q}^{k} + d_{\omega} d_{Q}^{\omega},
\end{align}
where $ d_{Q}^{k} $ and $ d_{Q}^{\omega} $ are the momentum and frequency scaling dimensions, respectively. Parameter $ d_{\omega} $ is usually chosen according to 
 the assumed  dispersion law of the free theory $ \omega \sim k^{d_{\omega}} $, which in our case corresponds to $ d_{\omega} = 2 $. In 
 order to obtain a renormalizable model, we 
  have to eliminate all divergences arising in vertex functions \eqref{1:eq.G} with 
  a non-negative value of UV exponent
\begin{align}
  d_{\G} = d_c + 2 - \sum_{\Phi} d_{\Phi} n_{\Phi},
\end{align}
where $d_c$ is the upper critical dimension of the model (in the present case $d_c=4$),  $ d_{\Phi} $ is the total canonical dimension of the
 field $\Phi$, and $ n_{\Phi} $ is the total number of the field $ \Phi $
 appearing in a given function $\G$, all calculated at $d=d_c$. All canonical 
dimensions for the active fluid model can be found in Tab.~\ref{tab:canon_v}. The degree of divergence of any vertex function $\G$ for model \eqref{1:eq.SSR} 
 is then given by the formula
\begin{align}
\label{degree}
d_{\G} = 6 - N_{v} - 3N_{v'}.
\end{align}
It should be noted that due to appearance of closed loops of retarded propagators all non-vanishing one-irreducible functions must have at least one
 field argument $v'$, i.e. $N_{v'}\ge 1$. Taking this into account and relation (\ref{degree})
we find  that ultraviolet divergences are present in the following
one-irreducible functions 
\begin{align}
\G^{v'v}: & \quad \text{with counterterms} \quad v'\partial_{t}v, \ v'\partial^{2}v, \ \tau v'v, \\
\G^{v'vv}: & \quad \text{with counterterm} \quad v'(v\partial)v, \\
\G^{v'vvv}: & \quad \text{with counterterm} \quad v'vvv,\\
\G^{v'v'}: & \quad \text{with counterterm} \quad v'v', 
\end{align}

In order to eliminate all divergences it is sufficient to renormalize 
fields according to prescription
\begin{equation}
  v_{i} \rightarrow v_{i} Z_{v}, \quad v_{i} \rightarrow v_{i}' Z_{v'}, 
\end{equation}
and parameters of the model in the following way
\begin{align}
  g_{10} &= \mu^{\varepsilon} g_{1} Z_{g_{1}},  &g_{20}& = 
  \mu^{\varepsilon} g_{2} Z_{g_{2}},  &\lambda_{0}& = \mu^{\varepsilon/2} 
  \lambda   Z_{\lambda}, \\ 
  \tau &= Z_{\tau} \tau_{0} + \tau_{c},  &\nu& = \nu_{0} Z_{\nu}.
\end{align}
Here, $g_{20}=\lambda_{0}^2$ stands for the expansion parameter of the perturbation theory and $ \mu $ is the renormalization mass scale - an arbitrary parameter that appears in the renormalization process \cite{Vasilev04,Zinn}. Universal quantities and properties  are independent of $\mu$. The parameter $ \tau_{c} $ is required due to the additive renormalization of the mass parameter $ \tau_{0} $, but its corrections are not  captured within the dimensional renormalization method employed in this paper. The renormalized response functional finally takes the form
\begin{align}
	\s_{R}^{\text{SR}}[\vv',\vv] =&\ v_{i}' \Big\{ Z_{1} \partial_{t}
	  + \nu(Z_{2}(-\partial^{2}) + Z_{3}\tau) \Big\} v_{i} \nonumber \\
	  &\ + \nu v_{i}'
	  \Big( Z_{4} 
	  \lambda    \mu^{\varepsilon/2} v_{j} \partial_{j} + Z_{5} g_{1} \mu^{\varepsilon} |\vv|^{2} /3! \Big) v_{i},  \nonumber \\
	  &\ - Z_{6}\nu_{0} v_{i}' \PP_{ij} v_{j}'/2,
	  \label{eq:renorm_act}
\end{align}
which is augmented with the following renormalization of the kernel function  
\begin{equation}
   D_{ij}^{\vv}(\kk) = Z_{6}\nu \PP_{ij}(\kk).
   \label{eq:renorm_kernel}
\end{equation}
In Eqs.~\eqref{eq:renorm_act} and \eqref{eq:renorm_kernel} we have introduced the following short-hand notation for 
renormalization constants 
\begin{align}
  Z_{1} & = Z_{v}Z_{v'}, 
  &Z_{2}& = Z_{1} Z_{\nu}, 
  &Z_{3}& = Z_{2} Z_{\tau},
  \nonumber \\
  Z_{4} & = Z_{2} Z_{\lambda} Z_{v},
  &Z_{5}& = Z_{2} Z_{g_{1}} Z_{v}^{2}, 
  &Z_{6}& = Z_{\nu} Z_{v'}^{2}.
  \label{I:Eq.Z}
\end{align}
The inverse relations are readily found 
\begin{align}
  Z_{\lambda}& = Z_{1}^{-1/2} Z_{2}^{-3/2} Z_{4} Z_{6}^{1/2},  
  &Z_{\tau}& = Z_{3} Z_{2}^{-1},   
  \nonumber \\
  Z_{v} & = Z_{1}^{1/2} Z_{2}^{1/2} Z_{6}^{-1/2},
  &Z_{v'}& = Z_{1}^{1/2}Z_{2}^{-1/2}Z_{6}^{1/2}, 
  \nonumber\\  
  Z_{g_{1}} &= Z_{1}^{-1} Z_{2}^{-2} Z_{5} Z_{6}, 
  &Z_{\nu} & = Z_{2} Z_{1}^{-1},
  \nonumber\\  
  Z_{g_{2}}& = Z_{\lambda}^{2}.
  \label{I:Eq.Zinversed}
\end{align}
In order to renormalize the model to the one-loop order we have to analyze following expansion of vertex functions \eqref{1:eq.G}
\begin{align}
  \Gamma_{v_{i}'v_{j}}& =  \ i\Omega Z_{1} - \nu k^{2} Z_{2} - \nu \tau Z_{3} +  \raisebox{-0.2cm}{\includegraphics[width=1.75cm]{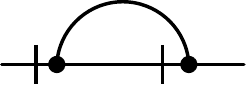}}
  \nonumber\\
  & + \frac{1}{2} \ \raisebox{-0.25cm}{\includegraphics[width=1.45cm]{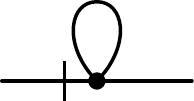}} , 
  \\
  \Gamma_{v_{i}'(\pp)v_{j}v_{k}}& =  \  i \lambda \mu^{\varepsilon/2} \nu Z_{4} 
  \mathbb{T}_{ijk}(\pp) + 
   \raisebox{-0.36cm}{\includegraphics[width=1.75cm]{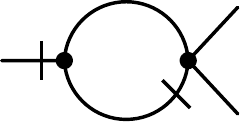}} 
    \nonumber\\ 
   &  +2 \ \raisebox{-0.33cm}{\includegraphics[width=1.7cm]{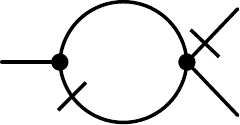}} 
   + \raisebox{-0.6cm}{\includegraphics[width=1.5cm]{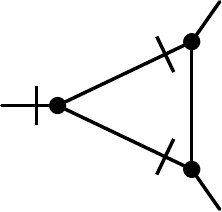}} 
   + 2 \ \raisebox{-0.6cm}{\includegraphics[width=1.5cm]{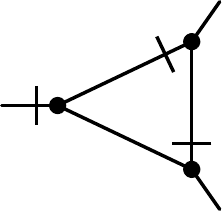}} , \nonumber \\
  \Gamma_{v_{i}'v_{j}v_{k}v_{l}} &= \  - g_{1} \mu^{\varepsilon} \nu  Z_{5}
  \mathbb{F}_{ijkl}  + 
  3 \ \raisebox{-0.35cm}{\includegraphics[width=1.65cm]{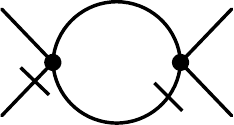}} 
  \nonumber \\
  & \ + 3 \ \raisebox{-0.65cm}{\includegraphics[width=1.5cm]{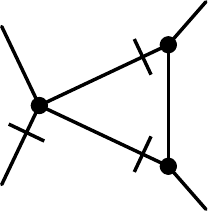}} 
  + 6 \ \raisebox{-0.65cm}{\includegraphics[width=1.5cm]{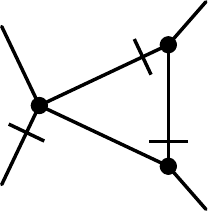}} , \label{1:eq.Gvsvvv} \\
  \Gamma_{v_{i}'v_{j}'} &= \  \nu Z_{6} \PP_{ij}(\kk) ,
\end{align}
with higher order corrections neglected. The approach for calculating Feynman diagrams and the resulting normalization constants can be found  in appendices \ref{app:FD} and \ref{app:RC}. Let us remark that there is no contribution to the renormalization of the random force correlator within the one-loop approximation as in the classical compressible Navier-Stokes equation in the vicinity of  $ d = 4 $ due to the transversality of the velocity field $ \vv $ \cite{AGK17}. However, this situation will change in two-loop approximation (where corrections appear due to presence of the quartic vertex), and therefore no exact relations between scaling exponents can be derived in general.

{\subsection{Critical scaling}
\label{subsec:rg}
}
The investigation of the large spatial and long temporal properties requires a thorough analysis of the Green functions at different scales~\cite{Vasilev04,Tauber,Zinn}. The 
 fundamental relation between renormalized and bare Green functions takes form
\begin{align}
  G_{0}^{(N_{v},N_{v'})}(\{\Bbbk_{i}\},e_{0}) =& \ Z_{v'}^{N_{v'}}(g) Z_{v}^{N_{v}}(g)
  \nonumber \\
 & \times G^{(N_{v},N_{v'})}(\{\Bbbk_{i}\},e,\mu), 
\end{align}
where $\{\Bbbk_i\} = \{ (\kk_i,\omega_i) \}_{i=1}^{N_v+N_{v'}}$ is meant as a short-hand for all external frequencies and momenta entering a Green function, $ e_{0} \equiv\ \{ g_{0},\nu_{0},\tau_{0} \} $ is the set of all bare parameter, $ e = e(\mu) $ are their renormalized counterparts at the scale $ \mu$, $ g = g(\mu) \equiv \{ g_{1},g_{2}\} $ is the set of all renormalized charges, $ N_{v}$ is the number of fields $ v $ entering the function $G_0$, and $N_{v'} $ is the number
of corresponding response fields $ v' $.

We define differential operators $ \D_{x} = x\partial_{x}|_{e} $ and $ \tilde{\D}_{x} = x\partial_{x}|_{e_{0}} $ to be logarithmic differential 
operators with respect to the renormalized parameters and bare parameters fixed, respectively. The investigation at different scales requires
 performing the logarithmic partial derivative
 with respect to the $ \mu $ while holding bare parameters $ e_{0} $ fixed. This yields fundamental RG equation
\begin{align}
\{ \D_{\mu} + \beta_{g} \partial_{g} -\gamma_{\nu} \D_{\nu} - \gamma_{\tau} & \D_{\tau} + N_{v} \gamma_{v} + N_{v'} \gamma_{v'} \} \nonumber \\
&\ \hspace{0.5cm} \times G(\{k_{i}\},e,\mu) = 0, \label{1:eq.RGEq}
\end{align}
with
\begin{align}
\beta_{g} = \tilde{\D}_{\mu} g, \quad \gamma_{x} = \tilde{\D}_{\mu} \ln Z_{x}, \label{1:eq.gBeta}
\end{align} 
being corresponding beta function $\beta_g$ of charge $g$ and anomalous
 dimension $\gamma_x$ of quantity $x$. We calculate the latter from the renormalization constants by means of an approximate relation
\begin{align}
\gamma_{x} &= (\beta_{g_{1}} \partial_{g_{1}} + \beta_{g_{2}} \partial_{g_{2}}) \ln Z_{x}, \\
&\simeq -\varepsilon (\D_{g_{1}} + \D_{g_{2}}) \ln Z_{x}.
\end{align}
Relations between various anomalous dimensions can be then found from the relations between the renormalization
 constants~\eqref{I:Eq.Zinversed}, from which we infer
\begin{align}
\gamma_{v} &= \frac{ \gamma_{1} + \gamma_{2} - \gamma_{6}}{ 2},  & \gamma_{v'}& =  \frac{\gamma_{1} - \gamma_{2} + \gamma_{6}}{2}, \label{1:eq.gamma1} \\
\gamma_{g_{1}}& =  \gamma_{5} + \gamma_{6}- \gamma_{1} - 2 \gamma_{2}, & \gamma_{g_{2}} & = 2 \gamma_{\lambda}, \\ 
\gamma_{\lambda} & = \frac{2 \gamma_{4}- \gamma_{1} - 3 \gamma_{2}  - \gamma_{6} }{2}, \\
\gamma_{\tau}& = \gamma_{3} - \gamma_{2},  &\gamma_{\nu} & = \gamma_{2} - \gamma_{1}. \label{1:eq.gamma2}
\end{align}
The explicit form of the anomalous dimensions is given in appendix \ref{app:RC}. The general form of beta functions is
 $ \beta_{g} = -g(d_{g} + \gamma_{g}) $ and for the present model we have
\begin{align}
\beta_{g_{1}} &= - g_{1} \left( \varepsilon - \frac{17 g_{1}}{24} - \frac{g_{2}}{4} \right), \\
\beta_{g_{2}} &= - g_{2} \left( \varepsilon - \frac{5 g_{1}}{18} - \frac{3 g_{2}}{8} \right).
\end{align}
At the fixed point $g^*$ both beta functions identically vanish, i.e. 
$\beta_{g_1}(g^*) = \beta_{g_2} (g^*) = 0$
 The stability of a given fixed point is determined by the
  eigenvalues $ \lambda $ of the matrix 
\begin{align}
\Omega_{ij} = \partial_{g_{i}} \beta_{g_{j}},
\end{align}
where $ \text{Re}[\lambda] > 0 $ for a stable fixed point. All fixed points and the corresponding eigenvalues 
of stability matrix are listed in Tab. \ref{tab:FP_SR}.

\begin{table}[t!]
	\def\arraystretch{1.5}
	\centering
	\begin{tabular}{| c || c | c || c | c |}
		\hline
		FP/$ g_{i}^{*} | \lambda_{i} $ & $g_{1}^{*}$ & $g_{2}^{*}$ & $\lambda_{1}$ & $ \lambda_{2} $ 
		\\  \hline \hline
		FP0 - Gaussian & $0$ & $0$ & $-\varepsilon$ & $-\varepsilon$
		\\  \hline
		FPI - Navier-Stokes & $0$ & $\frac{8}{3}\varepsilon$ & $-\frac{1}{3}\varepsilon$ & $\varepsilon$ 
		\\  \hline
		FPII - Model A & $\frac{24}{17}\varepsilon$ & $ 0 $ & $\varepsilon$ & $-\frac{31}{51}\varepsilon$
		\\ \hline 
		FPIII - Active Fluid & $\frac{72}{113}\varepsilon$ & $\frac{248}{113}\varepsilon$ & $\varepsilon$ & $ \frac{31}{113}\varepsilon $  
		\\ \hline    
	\end{tabular}
	\caption{Fixed points with their regions of stability.}
	\label{tab:FP_SR}
\end{table}

The first fixed point FP0 represents the Gaussian fixed point, for which all
 interactions 
  are infrared irrelevant. This fixed point is IR stable for $ \varepsilon < 0 $ or $ d > 4 $ as expected. The following two fixed points
   FPI and FPII represent the Navier-Stokes and the transversal Model A universality class. Both of these fixed points are unstable for 
  any value of the exponent $ \varepsilon $. The last fixed point FPIII represents a new universality class of incompressible 
  active fluid already reported recently in work~\cite{CTL15}.

In statistical physics, we are mainly interested in the macroscopic behavior of the
two-point correlation function
\begin{equation}
G^{(2,0)}(\kk,\omega) = k^{2\Delta_v-d-\Delta_\omega} G_{\pm}^{(2,0)}
\left( \frac{\tau}{k^{\Delta_\tau}}, \frac{\omega}{ \nu k^{\Delta_\omega}}
\right),
\end{equation}
where $ \pm $ denotes behavior above and below the phase transition, respectively. The critical exponents 
$ \eta,1/\nu $ and $ z $, are found by solving the differential RG equation \eqref{1:eq.RGEq} at the fixed point. This can be achieved in 
a straightforward manner~\cite{Vasilev04,Tauber}, and in principle the general 
	scaling formula for connected Green function
	can be written down.
Let us note that the total scaling dimension of any quantity $ Q $ is introduced as 
follows~\cite{Vasilev04}
\begin{align}
\Delta_{Q} = d_{Q}^{k} + \Delta_{\omega} d_{Q}^{\omega} + \gamma_{Q}^{*}, \quad \Delta_{\omega} = 2 - \gamma_{\nu}^{*}, \label{1:eq.critExp1}
\end{align}
with $ d_{Q}^{k} $ and $ d_{Q}^{\omega} $ being the momentum and frequency canonical dimensions from Tab.~\ref{tab:canon_v}. 
Critical exponents are then traditionally defined as 
\begin{equation}
\eta = 2 - d + 2 \Delta_{v}, \quad 1/\nu = \Delta_{\tau}, \quad z = \Delta_{\omega}. \label{1:eq.critExp2}
\end{equation}
For the FPIII we thus obtain the following prediction for critical exponents 
\begin{align}
\eta = \frac{31}{113} \varepsilon, \quad 1/\nu = 2 - \frac{58}{113} \varepsilon, \quad z = 2 - \frac{31}{113} \varepsilon, 
\end{align}
which is in agreement with the original results by Chen \emph{et al.} \cite{CTL15}.

{\section{The effect of non-local shear stress} \label{sec:nonlocal}}
{\subsection{Mesoscopic description} \label{subsec:meso}}

In this section we assume that shear stress of the active fluid is non-local in nature, i.e. active particles can feel "long-range friction forces". This leads to superdiffusive properties of the velocity field, which can be also viewed as the consequence of active particles performing L\'evy flights~\cite{CL19a,CL19b}. As a result, non-local shear stress increases kinetic energy dissipation, which reduces the correlations in the system. We will first formulate the hydrodynamic model, and leave the detailed discussion about its relation to microscopic features of the matter at the end of this section. 

In order to study non-local effects, we modify the strain rate tensor~\cite{evans_book} in the following way
\begin{align}
\varepsilon_{ij}(\xx) \rightarrow \eps_{ij}(\xx) + \int \dd^{d} y \ \eps_{ij}(\yy) \kappa(|\yy-\xx|). \label{Eq.stress1}
\end{align}
The first term on the right hand side represents the classical local strain stress, and the second term represents the 
non-local contributions. The kernel function $ \kappa(|\rr|) $ weights the contributions from the long-distance points 
and by taking into account the isotropy of the system, we expect it to follow an ordinary power law
 $ \kappa(|\rr|) \sim 1/|\rr|^{d-2\alpha} $. Using the definition of 
the Riesz fractional integro-differential formula \cite{SKM93}
\begin{align}
[\mathrm{I}^{2\alpha} \eps_{ij}](\xx) &\equiv \int \frac{\dd^{d} y}{C_{2\alpha}} \ \frac{\eps_{ij}(\yy)}{|\xx-\yy|^{d-2\alpha}} 
\rightarrow (-\partial_{\xx}^{2})^{-\alpha}, \label{eq:nonLocStress} \\
C_{\alpha} &= 2^{\alpha} \pi^{d/2} \frac{\G(\alpha/2)}{\G((d-\alpha)/2)}, 
\end{align}
where $ \G(\ldots) $ is the Gamma function, the Toner-Tu model with long-range interactions takes the following form
\begin{align}
\partial_{t} v_{i} + \lambda_{0} v_{j}\partial_{j} v_{i} & = \nu_{0} \partial^{2} v_{i} - w_{0}
 ( -\partial^{2})^{1-\alpha} v_{i} - \partial_{i} \tilde{p} \nonumber \\
& - \left(\tau_{0} + g_{10}|\vv|^{2} / 3! \right)
 v_{j} + f_{i}, \label{eq:non-locTT} \\
\partial_{i} v_{i} & = \ 0, 
\end{align}
where $ w_{0} $ is a mesoscopic amplitude. For the completeness we also mention the vorticity equation for the non-local Toner-Tu theory 
\begin{align}
\partial_{t} \omega_{i} & =\  \nu_{0} \partial^{2} \omega_{i}  -w_{0}(-\partial^{2})^{1-\alpha} \omega_{i} \nonumber \\
& - \tau_{0} \omega_{i} + f_{i} + \mathcal{O}(v^{2}) \label{eq:non-locTTvort}
\end{align}  
Eqs. \eqref{eq:non-locTT} and \eqref{eq:non-locTTvort} imply that both velocity and vorticity field undergo fractional diffusion process, which
 implies that any perturbation to the velocity and vorticity field will smooth out much faster that in the local case $ \alpha = 0 $. This indicates
  much weaker correlations in the present non-local theory, which can be already seen from the real-space representation of the equal-time correlation function
 (in the limit $ \nu_{0} = \tau_{0} = 0 $)
\begin{align}
  \langle v_{i}(\xx,t) v_{j}(\yy,t) \rangle_{0} \propto \frac{1}{|\xx-\yy|^{d-2(1-\alpha)}},
\end{align}
The origin of this phenomenon can be found in the energy balance equation \eqref{1:eq.TTc}, which is in the case with LR interactions modified as 
\begin{align}
\partial_{t} E & =  - 2\nu_{0} \langle\langle \omega^{2} \rangle\rangle - 2 w_{0} \langle\langle \omega_{i} (-\partial^{2})^{-\alpha}
 \omega_{i} \rangle\rangle - \nonumber \\
& - 2 \tau_{0} E - \frac{1}{3!} g_{0} \langle\langle v^{4} \rangle\rangle + \langle\langle f_{i}v_{i} \rangle\rangle, \label{Eq:10a}
\end{align}
where 
\begin{align}
\langle\langle \omega_{i} (-\partial^{2})^{-\alpha} \omega_{i} \rangle\rangle (t) = 
\int \dd^{d} x \dd^{d} y \ \frac{\omega_{i}(\xx,t) \omega_{i}(\yy,t)}{|\xx-\yy|^{d-2\alpha}}. 
\end{align}
Eq. \eqref{Eq:10a} implies larger energy dissipation, where the additional losses are caused by the LR interactions of the vorticity vector with  the weight function determined by the exponent $ \alpha $. In addition, the parameter $ w_{0} $ can be interpreted as a "long-range viscosity" in this system. At this point it is natural the expect, that for the values of $ \alpha $ high enough, any kind of perturbation will be immediately smooth out and this type of decorrelation will eventually lead to the mean-field description. These predictions are confirmed by the RG analysis below. 

Let us now discuss how to relate the non-local shear stress \eqref{eq:nonLocStress} with the microscopic features of the matter. As mentioned in the introduction, one way how to think about it is that it describes the ability of active particles to sense perturbations in the velocity field over longer distances. Moreover, recent experiments on starling flocks show, that the interactions between individual particles are ruled by the topological rather than metric distances \cite{Ballerini2008}. One may speculate, that this type of "long-range" interactions may be effectively captured by the expression \eqref{eq:nonLocStress}, where the precise value of $ \alpha $  would have to be determined from the statistical properties of inter-particle distances measured in an experiment.
	
The more experimentally appealing interpretation of the non-local shear stress is the result of active particles performing L\'evy flights. However, It should be noted that our model \eqref{eq:non-locTT} differs from the "Active L\'evy Matter" model derived by the means of the Boltzmann's approach due to the absence of certain non-local terms  \cite{CL19a,CL19b}. In the present work, the mesoscopic hydrodynamic equation \eqref{eq:non-locTT} has been derived using solely phenomenological ideas in order to capture the superdiffusive properties of the velocity field, and to keep the RG analysis simple. In order to quantify the relation between the non-local shear stress \eqref{eq:non-locTT} and "L\'evy properties" of active fluids, the relation between the parameter $ \alpha $ and dynamical properties of individual particles is needed.
	
The effect of anomalous dispersion on diffusion is, however, a complex problem of itself and requires modelling of transport of a scalar field in the velocity field, which we have not done explicitly. A similar problem has been studied in detail in the case of non-Galilean invariant 'synthetic' velocity field with anomalous dispersion law \cite{A99Synthetic,A00,Adzhemyan2002}, where it has been found that it may change diffusion of the scalar quantities. As one may expect, the solenoidal part of the velocity field enhances diffusion, whereas the potential part tends to slow it down. 
	
Since the current model \eqref{eq:non-locTT} is purely solenoidal, we expect that the non-local shear stress \eqref{eq:nonLocStress} is related to the enhanced diffusivity of the particles. The non-trivial calculation of the exact quantitative relation for all values of $ \alpha $ is, however, beyond the scope of this work, which we hope to carry out in the near future.

{\subsection{Field-theoretic renormalization } \label{subsec:nonlocal_rg}}
De Dominicis-Janssen response functional for the incompressible active fluid with LR interactions has the following form
\begin{align}
\label{DDJMixed}
\s^{\text{LR}}[\bar{\vv}',\bar{\vv}] &=  \bar{v}_{i}' \Big\{ \partial_{t} + \bar{\nu}_{0}(-\partial^{2} + \bar{w}_{0}(-\partial^{2})^{1-\alpha} + \bar{\tau}_{0}) \Big\} \bar{v}_{i} \nonumber \\
& + \bar{\nu}_{0} \bar{v}_{i}'
\Big( \bar{\lambda}_{0} \bar{v}_{j} \partial_{j}  + \bar{g}_{10} |\bar{\vv}|^{2}/3! \Big) \bar{v}_{i}  \nonumber \\
& - \bar{\nu}_{0} \bar{v}_{i}' \PP_{ij} \bar{v}_{j}'/2,
\end{align}
where we have again rescaled parameters with viscosity due to the dimensional reasons (with $ w_{0} \rightarrow \bar{\nu}_{0} \bar{w}_{0} $) and 
we have relabeled all parameters with bar in order to distinguish parameters with their short-range counterparts. Note that the non-locality 
appears only in the quadratic part of the response functional, while the interaction terms remain unchanged.

In the following considerations we may apply the same field-theoretic methods as have been used in the previous Sec.~\ref{sec:class}. There are, however, certain 
important issues that are needed to be discussed. Let us note that for $ \alpha > 0 $, the SR diffusion term is IR 
irrelevant in comparison to the LR diffusion. One might think that the SR term can be neglected in our analysis, i.e. study
 only the limit $ \bar{w}_{0} \rightarrow \infty $.  This approach may however lead to discontinuities in critical
  exponents between $ \alpha < 0  $ and $ \alpha > 0 $ regimes \cite{Janssen2008}.
  The point is that even {\color{red}if} the SR term in initially discarded, it is generated by renormalization and acquires anomalous scaling dimension of its own. Therefore, the total  scaling dimension of the SR term may well be less than the canonical dimension $2$.
 In order to resolve this problem, one
   has to study the model in the region in which the Fisher exponent of the SR model
   (given by the anomalous dimension of the basic field: $\eta=2\gamma_v^*$)
   is of the same order as that of the LR model (the latter has a fixed value $\eta_{LR}=2\alpha$), i.e. consider the model with both SR and LR
    interactions with $ \alpha \sim \gamma_v^*$ \cite{HN89}. 
The model is renormalized in the sense of the double-expansion scheme, where from the practical point of view
 the ray scheme $ \alpha \propto \varepsilon $ is often employed~\cite{AHK05}. 

The Feynman diagrammatic structure discussed in Sec.~\ref{sec:class} does not exhibit substantial differences of topological structure. However, the 
propagators
 attain the following form
\begin{align} 
\langle \bar{v}_{i} \bar{v}_{j}' \rangle_{0} (\kk,\omega) &= \frac{\mathbb{P}_{ij}}{-i\omega + \bar{\nu}_{0} (k^{2} + \bar{w}_{0} k^{2(1-\alpha)} + \bar{\tau}_{0})}, \label{II:prop1} \\
\langle \bar{v}_{i} \bar{v}_{j} \rangle_{0} (\kk,\omega) &= \frac{\bar{\nu}_{0}\mathbb{P}_{ij}}{\omega^{2} + \bar{\nu}_{0}^{2}(k^{2}  + \bar{w}_{0} k^{2(1-\alpha)} + \bar{\tau}_{0})^{2}}. \label{II:prop2}
\end{align}

\begin{table}[t!]
	\def\arraystretch{1.5}
	\centering
	\begin{tabular}{| c | c | c | c | c | c | c | c | c |}
		\hline
		$Q$ & $\bar{{\bm v}}$ & $\bar{{\bm v}}'$ & $\sqrt{\bar{\tau}},\mu,\Lambda$ & $\bar{\nu}_0,\bar{\nu} $ & $\bar{g}_{10},\bar{g}_{20}=\bar{\lambda}_{0}^{2}$ & $\bar{w}_{0} $ & $\bar{w},\bar{\lambda},\bar{g}_{1},\bar{g}_{2}$
		\\  \hline
		$d_Q^k$ & $\frac{d}{2}-1$ & $\frac{d}{2}+1$ & $1$ & $-2$ & $\varepsilon$ & $\alpha$ & $ 0 $
		\\  \hline
		$d^\omega_Q$ & $0$ & $0$ & $0$ & $1$ & $0$ & $0$ & $ 0 $
		\\  \hline
		$d_Q$ & $\frac{d}{2}-1$ & $\frac{d}{2}+1$ & $1$ & $0$ & $\varepsilon$ & $\alpha$ & $ 0 $
		\\ \hline    
	\end{tabular}
	\caption{Canonical dimensions of the bare fields and bare parameters for the active fluid model with non-local shear stress.}
	\label{tab:canon_d}
\end{table}
In order to obtain the large-scale properties, we have to again analyze canonical dimensions of the model.
There is, however, a delicate point to deal with. The very notion of unambiguous canonical dimensions relies on the generalized homogeneity of the propagators under simultaneous scaling of the frequency and wave number. This property is absent in (\ref{II:prop1}) and (\ref{II:prop2}). To fix canonical dimensions the 
quadratic term in the response functional (\ref{DDJMixed})
corresponding to one of the wave-number dependent terms in the propagator denominators may be treated as a part of the interaction. Practically it is convenient to include the non-local term in the interaction part so that, as it can be seen from the list of canonical dimensions listed in Tab.~\ref{tab:canon_d}, the model has the same canonical dimensions as its SR counterpart 
 (with additional parameter $ \bar{w}_{0} $).
 
 At this point it is instructive to recall the degree of divergence in its generic form
 \begin{align}
 \label{degreeMixed}
  d_{\G} & = d + 2 -2\alpha V_2+\left({d\over 2}-2\right)V_3+\left(d-4\right)V_4\\
  & -\left({d\over 2}-1\right)N_v-\left({d\over 2}+1\right)N_{v'}.
\end{align}
Here, $V_2$, $V_3$ and $V_4$ are the numbers of two-, three- and four-point vertices, respectively, in the one-irreducible graph. It is seen that coefficients of all $V_i$'s vanish, when $d=4$ and $\alpha=0$ giving rise to a logarithmic model with natural UV-regulators $\varepsilon=4-d$ and $\alpha$ in a analytic-dimensional renormalization scheme.

Since the non-local terms do not renormalize, relation (\ref{degreeMixed}) leads to 
 the same divergent vertex functions as in the SR model. In order to eliminate these divergences, we renormalize all parameters of the theory as follows
\begin{align}
  \bar{g}_{1} &= \mu^{\varepsilon} \bar{g}_{10} \bar{Z}_{\bar{g}_{1}}, \!\!
  &\bar{g}_{20}& = \mu^{\varepsilon} \bar{g}_{2} \bar{Z}_{\bar{g}_{2}}, & \bar{w} &= \mu^{2\alpha} \bar{w}_{0} \bar{Z}_{\bar{w}}, \\ 
  \bar{\lambda} & =  \mu^{\varepsilon/2} \bar{\lambda}_{0} \bar{Z}_{\bar{\lambda}}
 ,\!\! &\bar{\tau}& = \bar{\tau}_{0} \bar{Z}_{\bar{\tau}} + 
  \bar{\tau}_{c} , & \bar{\nu} & = \bar{\nu}_{0} \bar{Z}_{\bar{\nu}},
\end{align}
with $ \bar{v}' \rightarrow \bar{v}' \bar{Z}_{\bar{v}'}, \ \bar{v} \rightarrow \bar{v} \bar{Z}_{\bar{v}} $. The renormalized action functional takes then form
\begin{align}
  \s_{R}^\text{LR}[\bar{\vv}',\bar{\vv}] &=   \bar{v}_{i}' \Big\{ \bar{Z}_{1} \partial_{t}     
  + \bar{\nu}(\bar{Z}_{2} (-\partial)^{2}
  \nonumber\\	
  & + \bar{w} \mu^{2\alpha}(-\partial^{2})^{1-\alpha} + \bar{Z}_{3} \bar{\tau}) \Big\} \bar{v}_{i}  \nonumber \\
  & + \bar{\nu} \bar{v}_{i}'
  \Big(  \bar{Z}_{4} \bar{\lambda} \mu^{\varepsilon/2} v_{j} \partial_{j}\! + \bar{Z}_{5}
  \bar{g}_{1} \mu^{\varepsilon} |\bar{v}|^{2} /3! 
  \Big) \bar{v}_{i} \nonumber \\
  & - \bar{Z}_{6} \bar{\nu} \bar{v}_{i}' \PP_{ij} \bar{v}_{j}'/2,
  \label{eq:LRaction}
\end{align}
	where the short-hand notation for the renormalization constants with bar above is the same as in \eqref{I:Eq.Z}-\eqref{I:Eq.Zinversed}. 
According to the general RG considerations~\cite{Vasilev04} the non-local term in the action~\eqref{eq:LRaction} is not  renormalized, which results
 in the 	
	relation $\bar{Z}_{\bar{w}} = \bar{Z}_{2}^{-1}$. Following vertex functions require renormalization
\begin{align}
	\G_{\bar{v}_{i}'\bar{v}_{j}} =&\ i\Omega \bar{Z}_{1} - \bar{\nu} k^{2} \bar{Z}_{2} - \bar{\nu} \mu^{2\alpha} \bar{w} k^{2(1-\alpha)} \nonumber \\
	&\ - \bar{\nu} \bar{\tau} Z_{3} + \dots, \\
	\G_{\bar{v}_{i}'(\pp)\bar{v}_{j}\bar{v}_{k}} =&\ i \bar{\lambda} \mu^{\varepsilon/2} \bar{\nu} \bar{Z}_{4} \mathbb{T}_{ijk}(\pp) + \dots,  \\
	\G_{\bar{v}_{i}'\bar{v}_{j}\bar{v}_{k}\bar{v}_{l}} =&\ - \bar{g}_{1} \mu^{\varepsilon}  \bar{\nu} \bar{Z}_{5} \FF_{ijkl}  + \dots,  \\
	\G_{\bar{v}_{i}'\bar{v}_{j}'} =&\ \bar{\nu} \bar{Z}_{6} P_{ij}(\kk) + \dots,
\end{align}
and diagrammatic corrections display analogous perturbative structure as has been found in the SR model, but with 
propagators \eqref{II:prop1} and \eqref{II:prop2} and all quantities $ Q $ replaced with the LR quantities $ \bar{Q} $ (with bar).
{ \subsection{Asymptotic behavior} \label{subsec:asymbeh} }
As in the case of the model with SR interactions, we need to investigate the properties of the Green's functions at different scales
\begin{align}
G_{0}(\{k_{i}\},\bar{e}_{0}) =&\ Z_{\bar{v}'}^{N_{\bar{v}'}}(\bar{g}) Z_{\bar{v}}^{N_{\bar{v}}}(\bar{g}) G(\{k_{i}\},\bar{e},\mu),
\end{align}
where now we have $ \bar{e} = \{ \bar{g},\bar{\nu},\bar{\tau} \}, \ \bar{g} = \{\bar{g}_{1},\bar{g}_{2},\bar{w}\} $. Relations between anomalous dimensions if the LR model are the same as in the case of SR model \eqref{1:eq.gamma1}-\eqref{1:eq.gamma2}, with 
additional constraint $ \gamma_{\bar{w}} = -\gamma_{2} $. Beta functions then follow in a straightforward manner
\begin{align}
\beta_{\bar{g}_{1}} =&\ -\bar{g}_{1} \left( \varepsilon - \frac{17 \bar{g}_{1}}{24(1+\bar{w})^{2}} - \frac{\bar{g}_{2}}{4(1+\bar{w})^{2}} \right), 
\label{LRbetaW1}
\\
\beta_{\bar{g}_{2}} =&\ - \bar{g}_{2} \left( \varepsilon - \frac{5 \bar{g}_{1}}{18 (1 + \bar{w})^{2}} - \frac{3 \bar{g}_{2}}{8 (1 + \bar{w})^{2}} \right), 
\label{LRbetaW2}
\\
\beta_{\bar{w}} =&\ - \bar{w} \left( 2\alpha - \frac{\bar{g}_{2}}{8(1 + \bar{w})^{2}} \right), \label{LRbetaW3}
\end{align}
where $ \beta_{\bar{g}} = 0 $ at the fixed point, and the stability is determined by the eigenvalues $ \lambda $ of the matrix 
\begin{align}
\Omega_{ij} = \partial_{\bar{g}_{i}} \beta_{\bar{g}_{j}},
\end{align}
where $ \text{Re}[\lambda] > 0 $ for a stable fixed point. 

It turns out that the actual expansion parameters of the perturbation theory are 
\begin{equation}
  \bar{g}_{i}' =  \frac{\bar{g}_{i}}{ (1 + \bar{w})^{2} }, \ i=1,2.
\end{equation}
 The transformation into these new variables is not necessary in order to obtain results for finite $ \bar{w} $. 
 However, the above set of equations \eqref{LRbetaW1}-\eqref{LRbetaW3} does not possess a fixed point with nonzero value of $ \bar{w}^{*} $. Therefore, all
  fixed points belong to the universality classes already found in the model with SR interactions - SR Gaussian, SR Navier-Stokes, SR Model A and SR Active
   fluid models. However, as seen from Table \ref{tab:FP}, all SR fixed points are unstable to the long-range interaction, when the LR Fisher exponent is
    larger than the SR Fisher exponent: $2\alpha>2\gamma_v^*$. Therefore, an analysis of the model without the local term is called for.

 { \subsection{The pure long-range limit} }

In order to study the limit $ \bar{w} \rightarrow \infty $, i.e.~to study the pure LR limit (PLR), we perform the following substitution
\begin{align}
  \tilde{v}' & = \bar{v}' / \bar{w}_{0}^{1/2}, 
  & \tilde{v} & = \bar{v} \bar{w}_{0}^{1/2},  
  & \tilde{\nu}_{0} &= \bar{\nu}_{0} \bar{w}_{0}, \\ 
  \tilde{\tau}_{0}& = \bar{\tau}_{0} / \bar{w}_{0},
  & \tilde{g}_{10} & = \bar{g}_{10} /   \bar{w}_{0}^{2}, 
  &\tilde{\lambda}_{0} & = \bar{\lambda}_{0} / \bar{w}_{0}^{3/2}, \\
   \tilde{g}_{20} & = \bar{g}_{20} / \bar{w}_{0}^{3},
\end{align}
and the bare response functional takes the following form
\begin{align}
	\s[\tilde{\vv}',\tilde{\vv}] &= \tilde{v}_{i}' \Big\{ \partial_{t} +
	 \tilde{\nu}_{0}((-\partial^{2})^{1-\alpha} + \tilde{\tau}_{0}) \Big\} \tilde{v}_{i}
	 \nonumber\\
	 & + \tilde{\nu}_{0} \tilde{v}_{i}'
	 \Big( \tilde{\lambda}_{0}
	  \tilde{v}_{j} \partial_{j} + \tilde{g}_{10} |\tilde{\vv}|^{2} /3! 
	  \Big)
	   \tilde{v}_{i}, \nonumber \\
	   & - \tilde{\nu}_{0} \tilde{v}_{i}' \PP_{ij} 
	   \tilde{v}_{j}'/2 .
	   \label{eq:rescaled_func}
\end{align}
An important remark is now in order. Since the original LR coupling constant scales as $ \bar{w}_{0} \sim \Lambda^{2\alpha} $, the parameters
 and fields in the above rescaled response functional~\eqref{eq:rescaled_func} attain different canonical dimensions - see Tab.~\ref{tab:canon_TLR}. Note
 that from the rescaled response functional
 we deduce following relation 
\begin{equation} 
  d_{\omega} = 2(1-\alpha) .
\end{equation} 
 Another relevant fact is that parameters $ \tilde{g}_{10} $ and $ \tilde{g}_{20} $ have no longer the same
 canonical dimensions, as can be readily seen from the expression for the degree of divergence
\begin{align}
 \label{degreeLR}
  d_{\G} & = d + 2(1-\alpha) +\left({d\over 2}-2+3\alpha\right)V_3+\left(d-4+4\alpha\right)V_4 \nonumber \\
 & -\left({d\over 2}-1+\alpha\right)N_v-\left({d\over 2}+1-\alpha\right)N_{v'}.
\end{align}
The upper critical dimension is determined by putting the coefficient of $V_4$ equal to zero, which leaves $V_3$ with the positive coefficient $\alpha$. This means that graphs with the three-point interaction are IR-irrelevant and are thus discarded in the subsequent RG analysis. With only the four-point interaction left we arrive at a modification (due to the incompressibility condition) of the dynamical model A with a long-range interactions.

The degree of divergence at the upper critical dimension $d_c=4(1-\alpha)$
  \begin{equation}
 \label{degreeLRlog}
  d_{\G} = 6(1-\alpha) 
  -\left(1-\alpha\right)N_v-3\left(1-\alpha\right)N_{v'},
\end{equation}
contains a real parameter $\alpha$, therefore only divergences with zero degree of divergence affect the asymptotic behavior. This condition is  directly met by vertex functions $\G^{v'v'}$ and $\G^{v'vvv}$. The function $\G^{v'v}$ has the degree $2(1-\alpha)$, but it should be noted that the derivatives of
  this function with respect to frequency and the temperature parameter $\tau$ possess a zero degree of divergence and thus give rise to nontrivial renormalization.

{ 
In what follows we will skip most of the details, as they are analogous to the SR and LR case discussed above. Based on the UV analysis above, the one-loop corrections to the vertex functions have in the present case much simple form
\begin{align}
\Gamma_{\tilde{v}_{i}'\tilde{v}_{j}}& =  \ i\Omega Z_{1} - \tilde{\nu} k^{2(1-\alpha)} - \tilde{\nu} \tilde{\tau} Z_{3} + \frac{1}{2} \ \raisebox{-0.25cm}{\includegraphics[width=1.45cm]{vsv2}},
\\
\Gamma_{\tilde{v}_{i}' \tilde{v}_{j} \tilde{v}_{k} \tilde{v}_{l}} &= \  - \tilde{g}_{1} \mu^{\varepsilon-4\alpha} \tilde{\nu} Z_{5}
\mathbb{F}_{ijkl}  + 
3 \ \raisebox{-0.35cm}{\includegraphics[width=1.65cm]{vsvvv1}} , \\
\Gamma_{\tilde{v}_{i}'\tilde{v}_{j}'} &= \  \tilde{\nu} Z_{6} \PP_{ij}(\kk), \label{1:eqLR.Gvsvvv}
\end{align}
with the renormalization constants
\begin{align}
Z_{1} & = Z_{\tilde{v}}Z_{\tilde{v}'}, 
&Z_{3}& = Z_{1} Z_{\tilde{\tau}},
\nonumber \\
Z_{5}& = Z_{\tilde{g}_{1}} Z_{\tilde{v}}^{2}, 
&Z_{6}& = Z_{\tilde{v}'}^{2},
\label{I:Eq.ZW}
\end{align}
which can be found in Appendix \ref{app:RC}. }

\begin{table}[t!]
	\def\arraystretch{1.5}
	\centering
	\begin{tabular}{| c || c | c | c | c | c | c | c | c |}
		\hline
		$Q$ & $\tilde{{\bm v}}$ & $\tilde{{\bm v}}'$ & $\mu,\Lambda$ & $ \tilde{\tau}_{0},\tilde{\tau} $ & $ \tilde{\nu}_0,\tilde{\nu} $ & $ \tilde{g}_{10} $ & $ \tilde{g}_{20}=\tilde{\lambda}_{0}^{2}$ & $\tilde{\lambda},\tilde{g}_{1},\tilde{g}_{2} $
		\\  \hline \hline
		$d_Q^k$ & $\tfrac{d-z_{\alpha}}{2}$ & $\tfrac{d+z_{\alpha}}{2}$ & $1$ & $z_{\alpha}$ & $-z_{\alpha}$ & $ \varepsilon - 4\alpha $ & $ \varepsilon - 6\alpha $ & $ 0 $
		\\  \hline
		$d^\omega_Q$ & $0$ & $0$ & $0$ & $0$ & $1$ & $0$ & $0$ & $0$ 
		\\  \hline
		$d_Q$ & $\tfrac{d-z_{\alpha}}{2}$ & $\tfrac{d+z_{\alpha}}{2}$ & $1$ & $z_{\alpha}$ & $0$ & $ \varepsilon - 4\alpha $ & $ \varepsilon - 6\alpha $ & $ 0 $
		\\ \hline 
	\end{tabular}
	\caption{Canonical dimensions of the bare fields and bare parameters for AF model. Note that in this case $ z_{\alpha} = 2(1-\alpha) $. }
	\label{tab:canon_TLR}
\end{table} 

\begin{table*}
	\centering
	\def\arraystretch{1.5}
	\begin{tabular}{| c || c | c | c || c | c | c || c | c | c |}
		\hline
		FP/$ \bar{g}_{i}^{*} | \lambda_{i} $ & $\bar{g}_{1}^{*}|\tilde{g}_{1}^{*}$ & $\bar{g}_{2}^{*}|\tilde{g}_{2}^{*}$ & $ \bar{w}^{*} $ & $\lambda_{1}$ & $ \lambda_{2} $ & $ \lambda_{3} $  & $\eta$ & $1/\nu$ & $z$
		\\  \hline \hline
		FP0 - Gaussian & $ 0 $ & $ 0 $ & $ 0 $ & $ -\varepsilon $ & $ -\varepsilon $ & $ -2\alpha $ & $ 0 $ & $ 2 $ & $ 2 $ 
		\\  \hline
		FPI - SR Navier Stokes & $ 0 $ & $ \frac{8}{3}\varepsilon $ & $ 0 $ & $ -\frac{1}{3} \varepsilon $ & $ \varepsilon $ & $ \frac{1}{3} (\varepsilon - 6\alpha) $
		& $ \tfrac{1}{3} \varepsilon $ & $ 2 - \tfrac{1}{3}\varepsilon $ & $ 2 - \tfrac{1}{3}\varepsilon $ 
		\\  \hline
		FPII - SR Model A  & $ \frac{24}{17}\varepsilon $ & $ 0 $ & $ 0 $ & $ \varepsilon $ & $ -\frac{31}{51}\varepsilon $ & $ -2\alpha $
		& $ 0 $ & $ 2 - \tfrac{9}{17} \varepsilon $ & $ 2 $ 
		\\ \hline 
		FPIII - SR Active Fluid & $ \frac{72}{113}\varepsilon $ & $ \frac{248}{113}\varepsilon $ & $ 0 $ & $ \varepsilon $ & $  \frac{31}{113}\varepsilon $ & $ \frac{1}{113} (31 \varepsilon - 226\alpha) $  
		& $ \frac{31}{113}\varepsilon $ & $ 2 -\frac{58}{113}\varepsilon $ & $ 2 - \frac{31}{113}\varepsilon $ 
		\\ \hline \hline 
		FPIV - PLR Gaussian & $ 0 $ & $ 0 $ & $ \infty $ & $ -\tilde{\varepsilon} $ & ND & ND 
		& $ 2\alpha $ & $ 2(1-\alpha) $ & $ 2(1-\alpha) $ 
		\\  \hline
		FPV - PLR Model A & $ \frac{24(1-\alpha)(3-2\alpha)}{(3-4\alpha)(17-18\alpha+4\alpha^{2})} \tilde{\varepsilon} $ & $ 0 $ & $ \infty $  & $ \tilde{\varepsilon} $ & ND & ND
		 & $ 2\alpha $ &  $2(1 -\alpha)-A(\alpha)\tilde{\varepsilon} $  & $ 2(1-\alpha) $ 
		\\ \hline 
	\end{tabular}
	\caption{Fixed points with their stability eigenvalues, and the corresponding critical exponents. For convenience we have defined $ A(\alpha) = (3-2\alpha)^2/( 17 - 18\alpha + 4\alpha^{2}) $, and introduced following abbreviations SR - short-range, PLR - pure long-range. For pure long-range fixed points FPIV and FPV only one eigenvalue can be determined. Abbreviation then ND stands for Not Determined. Note, that the small expansion parameter for PLR fixed points is $ \tilde{\varepsilon} = 4(1-\alpha) - d $, rather than $ \varepsilon = 4 -d $, as in the case of SR fixed points.
    }	
	\label{tab:FP}
\end{table*}

The $\beta$ function for this model is
\begin{equation}
\label{betaLR}
\beta_{\tilde{g}_{1}} = - \tilde{g}_{1} \left( \tilde{\varepsilon} - \frac{(3 - 4 \alpha) \left(17 -18 \alpha + 4\alpha^{2}\right)}{24 (1-\alpha) (3 - 2 \alpha)} \tilde{g}_{1} \right), 
\end{equation}
It should be noted that the small parameter here is $\tilde{\varepsilon}=d_c-d=4-d-4\alpha=\varepsilon -4\alpha$, not $\varepsilon $ and $\alpha$ separately. 

\begin{figure}[t!]
	\includegraphics[width=8cm]{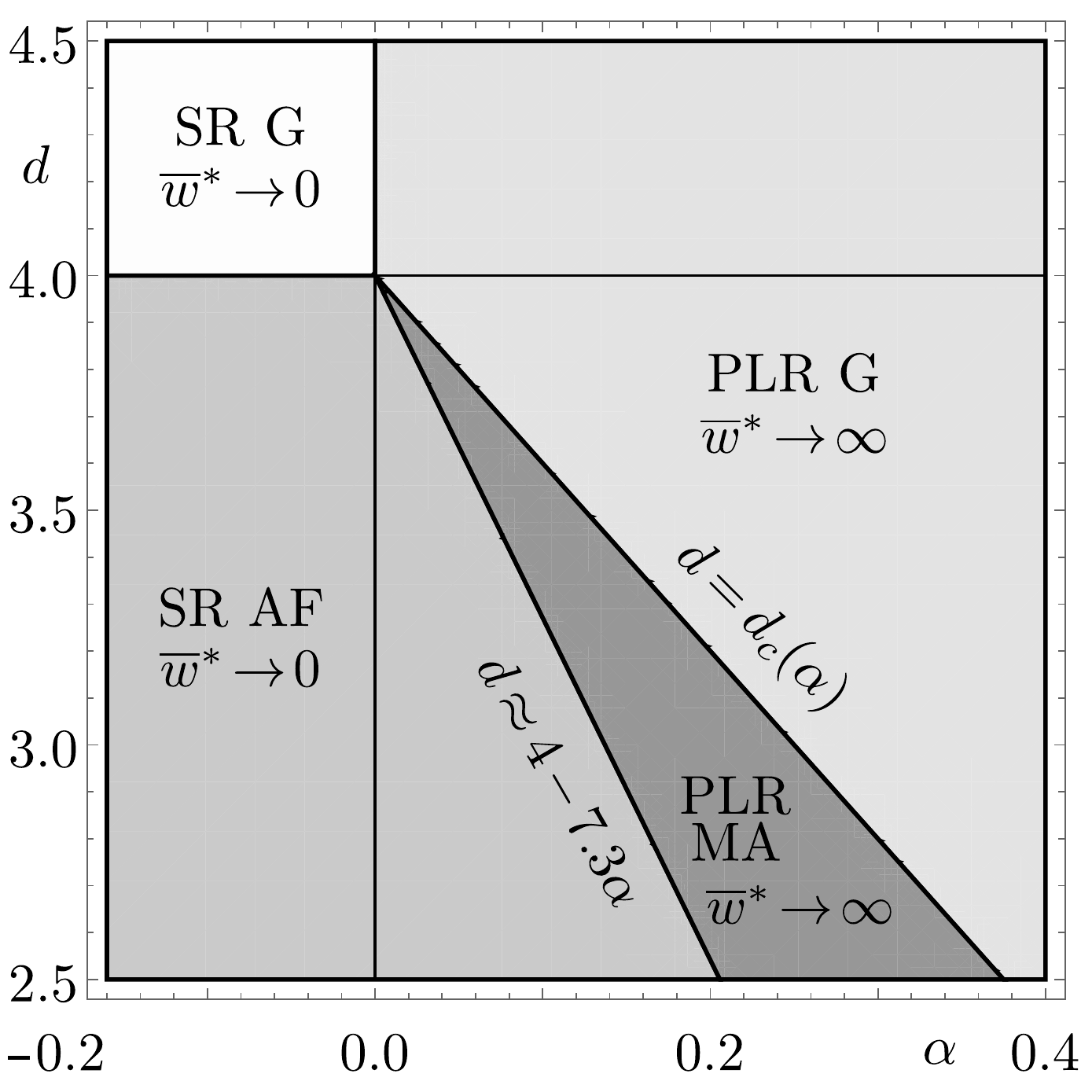}
	\caption{Phase diagram for different regimes. SR G Short range Gaussian fixed point. SR AF - Short range active fluid, PRL G - Pure Long Range Gaussian 
	fixed point and PLR MA - Pure Long Range Model A} \label{fig:phaseDiagram}
\end{figure}

The stability region of the single nontrivial fixed point (PLR Model A) is given by the condition $4-d-4\alpha>0$. The opposite inequality gives the stability condition of the Gaussian fixed point. The summary of all fixed points is reported in Tab.~\ref{tab:FP}. In this case, we have found only two new fixed points - PLR Gaussian
 and PLR Model A fixed point. It is interesting that in the one-loop approximation there is no PLR Active fluid nor PLR Navier-Stokes fixed point. As mentioned 
before, the absence of the latter might be due to lower scaling dimension of $ \tilde{g}_{1} $ than $ \tilde{g}_{2} $ for $ \alpha > 0 $. The
 physically most  realistic is
 the three-dimensional case $ \varepsilon = 1 (d=3) $. In this case for sufficiently small $ \alpha $ the model belongs to the universality class of
  SR Active fluid. However, at larger values
 $ \alpha > 31/226 \approx 0.137 $ there is a crossover to the PLR Model A universality class. Finally, for $ \alpha > 0.25 $ the effect
  of non-locality becomes so pronounced that
		all
  non-linearities become IR irrelevant and the mean-field approximation becomes valid. The phase diagram for this system is depicted in Fig.~\ref{fig:phaseDiagram}.

Critical exponents for all universality classes calculated according to \eqref{1:eq.critExp1}-\eqref{1:eq.critExp2} are shown in Tab.~\ref{tab:FP}. The
 overall analysis of fixed points FP0-FPIII is in accordance with the former results \cite{CTL15}, and although FPIV belongs to a 'standard' L\'evy
  diffusion universality class, the FPV represents a rather unusual 'long-range Model A' universality class, due to irrelevance of
   the short range 'diffusion' term and the incompressibility condition which remains valid even after the self-advection becomes
    irrelevant. To the best of our knowledge such fixed point has not been reported in the literature. One should also note, that
 at the boundary between the SR Active Fluid and PLR Model A regimes, i.e. for $ \alpha = 31\varepsilon/226 $, the independent 
 critical exponents $ \eta $, $ \nu $ and $ z $ change continuously for $ \varepsilon > 0 $.

{\section{Discussion and Conclusion} \label{sec:conclusion}}

In this work, we have been studying the effect of long-range interactions on the order-disorder phase transition in the incompressible dry active fluid. Starting from the incompressible Toner-Tu theory, we have extended the model by including non-local shear stress into the hydrodynamic description of the system, which led to appearance of the fractional viscous of a form $ \sim \partial^{2(1-\alpha)} $. This term is in general responsible for the non-local energy dissipation and leads to super-diffusive properties of the velocity field similar to the one observed in systems with L\'evy flights.  Using a standard approach, we have obtained the De Dominicis-Janssen response functional. The renormalization group procedure, based on the analysis of the UV divergences of the corresponding model was carried out. The analysis of the RG flow equations revealed six fixed points corresponding to six different universality classes. In the case of $ d = 3 $, we found that although for small values of $ \alpha $ the system belongs to the universality class of the incompressible active fluid, for $ \alpha \gtrsim 0.137 $ the self advection becomes irrelevant and a crossover to the universality class of the 'transversal Model A' with Long range (diffusion) interaction occurs. In addition, for $ \alpha > 0.25 $ the magnitude of the long-range interactions destroys the relevance of the non-linearities and the mean-field approximation becomes valid.

\subsection*{Acknowledgments}

The authors thank Alexander Morozov, Ervin Frey, and John Toner for many fruitful discussions. The authors are in debt to Chiu Fan Lee for bringing the works \cite{CL19a,CL19b} to their attention. 
V.~\v{S}kult\'ety acknowledges studentship funding from EPSRC grant no. EP/L015110/1.

The work was supported by VEGA grant No. 1/0345/17 of the Ministry of Education, Science, 
Research and Sport of the Slovak Republic.

\appendix

  \section{Calculation of Feynman diagram}   \label{app:FD}

  \subsection{Sunset diagram}
  For illustration purposes let us consider the sunset diagram
   \begin{equation}
     \frac{\PP_{12}(\pp)}{d-1} \times \ \raisebox{-0.4cm}{\includegraphics[width=3.2cm]{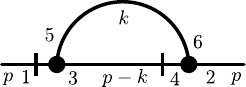}}, \label{app:diag_vsv1}
  \end{equation}
  which in the frequency-momentum representation corresponds to the algebraic expression
  \begin{align}
   \int\displaylimits_{\kk,\omega} \frac{-\bar{\lambda}_{0}^{2} \bar{\nu}_{0}^{3} \TT_{135}(\pp) \PP_{34}(\pp-\kk)\PP_{56}(\kk)\TT_{426}(\pp-\kk)}{[i(\omega-\Omega) 
  + \bar{\nu}_{0}(A_{\pp-\kk}+\bar{\tau}_{0})][\omega^{2} + \bar{\nu}_{0}^{2} (A_{\kk}+\bar{\tau}_{0})^{2}]},
  \label{eq:app_example}
  \end{align}
  where we have introduced the following abbreviations
\begin{align}
   \int\displaylimits_{\kk,\omega} & \equiv \frac{1}{(2\pi)^{d+1}} \int \dd^d k  \int \dd \omega, \\
   A_{\kk} & \equiv k^{2} + \bar{w}_{0} k^{2(1-\alpha)}.
\end{align}   

Let us note that for calculation purposes in expressions like~\eqref{eq:app_example} it is convenient to denote vector and tensorial indices by numbers rather than letters. Further, the summation over repeated indexes (numbers) in expression~\eqref{eq:app_example} is implied. Such abbreviation rule is always applied in this work. Since we are only interested in the transversal part
 of the diagram (the longitudinal part vanishes due to the transversality of the external propagators), we have contracted the diagram with 
 the expression $ \PP_{12}(\pp)/(d-1) $. Expanding the tensor structure we immediately observe that the diagram is already proportional to $ p $, so we can set external frequency $ \Omega = 0 $.   Performing the integration over internal frequency
 and keeping only terms proportional to $ p^{2} $ we  get
  \begin{multline}
 \frac{ -\bar{\nu}_{0}p^{2} \bar{g}_{20}}{4 d (d+2)} \Biggl\{
 \int\displaylimits_{\kk} \frac{ d^2-d-2}{(k^{2}  +
  \bar{w}_{0}k^{2(1-\alpha)}+\bar{\tau}_{0})^{2}}\\
  +
 2\int\displaylimits_{\kk} \frac{k^{2} + (1- \alpha)\bar{w}_{0}k^{2(1-\alpha)}}{(k^{2} + \bar{w}_{0}k^{2(1-\alpha)} + \bar{\tau}_{0})^{3}}\Biggr\},
  \end{multline}
  where we have used the following formulas \cite{Vasilev04} for tensor integrals
\begin{align}
  \int\displaylimits_{\kk} k_{1} k_{2} f(|\kk|^{2}) & = \frac{\delta_{12}}{d} \int_{\kk} k^{2} f(|\kk|^{2}), \label{app:diag_momInt1}\\
  \int\displaylimits_{\kk} k_{1} k_{2} k_{3} k_{4}  f(|\kk|^{2}) & = 
  \frac{\delta_{12}\delta_{34} +  (2 \ \text{perm.})}{d(d+2)} \int_{\kk} k^{4} f(|\kk|^{2}). \label{app:diag_momInt2}
\end{align}
  Now we can employ the 'Honkonen-Nalimov scheme'\cite{HN89}: one-loop counterterms are produced in the form of poles in $ \alpha,\varepsilon $ and their
   linear combinations multiplied by regular functions, therefore we can neglect any other dependence on these parameters in the sense that the residues 
   at poles generate the leading contribution to the RG functions $\gamma$ in $ \alpha$ and $\varepsilon $ in the one-loop approximation. The result is then 
\begin{align}
  - \frac{\bar{\nu}_{0}p^{2} \bar{g}_{20}}{8} \int\displaylimits_{\kk} 
  \frac{A_{\kk}}{(A_{\kk} + \bar{\tau}_{0})^{3}} \approx 
  -\frac{\bar{\nu}_{0}p^{2}\bar{g}_{20}}{8} \int\displaylimits_{\kk} \frac{1}{(A_{\kk} + 
  \bar{\tau}_{0})^{2}}, \label{app:diag_vsv1Cal}
\end{align}
  where the last step is valid because the integral is dominated by its UV behavior. Note that the choice of \eqref{app:diag_vsv1Cal} is justifiable since the
   integrals differ by a finite term, which is basically arbitrary (in two loops, however, finite terms of the one-loop subdiagrams have to be taken into 
   account just in the form chosen at the present stage in order to ensure multiplicative renormalization \cite{Vasilev04,Zinn}). In order to extract
    the singular part in parameters $ \alpha,\varepsilon $ of the resulting integral, we expand the denominator using the negative binomial series
\begin{align}
  \int\displaylimits_{\kk} \frac{\bar{g}_{20}}{(k^{2} + \bar{w}_{0}k^{2(1-\alpha)} + 
  \bar{\tau}_{0})^{2}} &= \sum_{l} \binom{-2}{l}
  \int\displaylimits_{\kk} \frac{\bar{g}_{20}\bar{w}_{0}^{l} k^{2l(1-\alpha)}}{(k^{2} + 
  \bar{\tau}_{0})^{2+l}} \nonumber \\
   &\hspace{-1.5cm} \approx \sum_{l} \binom{-2}{l} \frac{\bar{g}_{2} \bar{w}^{l}}{\varepsilon + 2l\alpha} 
  \left( \frac{\mu^{2}}{\bar{\tau}} \right)^{\frac{\varepsilon}{2} + l \alpha},
  \label{app:eq_deriv}
\end{align}
 where we have substituted $ S_{d}\bar{g}_{20}/(2\pi)^{d} \rightarrow \bar{g}_{20} $, with $ S_{d} $ being the surface of the $ d $-dimensional sphere.  We have  also expressed bare parameters in terms of renormalized ones, which is appropriate within the leading order.  The last integral in~\eqref{app:eq_deriv} was evaluated using the well known formula \cite{Vasilev04,Tauber}
  \begin{align}
  \int_{\kk} \frac{1}{(k^{2}+\tau_{0})^{a}k^{2b}} =& \nonumber \\
  & \hspace{-2.2cm} = \frac{S_{d}}{(2\pi)^{d}} \frac{\G(d/2-b)\G(a+b-d/2)}{2\G(a)} \tau_{0}^{d/2-a-b}, \label{app:Fint}
  \end{align}
  where $ \G(\ldots) $ stands for the Gamma function. The final form of the diagram \eqref{app:diag_vsv1} then reads
\begin{align}
  - \frac{\bar{\nu} p^{2}\bar{g}_{2}}{8} \sum_{l} \binom{-2}{l} 
  \frac{\bar{w}^{l}}{\varepsilon + 2l\alpha} \left( \frac{\mu^{2}}{\tau} 
  \right)^{\frac{\varepsilon}{2} + l \alpha}. 
  \label{App:eq.1}
\end{align}


  \subsection{Three point vertex}
  This three point vertex contribution is calculated for the case of SR interaction, the generalization to the LR system is straightforward.   
  We realize that every interaction vertex is to be contracted with three other projection operators (three connecting propagators),
\begin{align}
  \raisebox{-0.8cm}{\includegraphics[width=2.2cm]{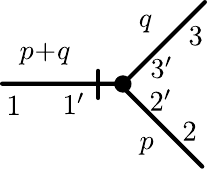}} & = i \lambda_{0} 
  \nu_{0}   \PP_{1'1}(\pp+\qq) \PP_{2'2}(-\pp) \PP_{3'3}(-\qq) 
  \nonumber\\ 
  & \times  
  \TT_{1'2'3'}(\pp+\qq), \label{App:eq.TV}
\end{align}
  where $ \TT_{1'2'3'} $ is a vertex tensor structure \eqref{1:eq.T1}.
   Expanding the above structure, we obtain the following tensor
    $ \mathbb{X}_{123}(\pp,\qq) $
\begin{align}
  \mathbb{X}_{123}(\pp,\qq) =&\ \PP_{34}(\qq) \left[ \delta_{12} p_{4} + 
  p_{1} p_{4}   \frac{ p^2 \left(p_2+2 q_2\right)+p_2 q^2 }{p^2 
  \left(\pp+\qq\right)^{2}} \right] 
 \nonumber \\   
 &\ + (2\leftrightarrow 3, 
  p \leftrightarrow q). \label{App:eq.TV2}
\end{align}
  This means, that in order to find vertex contributions we have to contract diagram results with three projection operators as  pointed out in \eqref{App:eq.TV} and the result has to be proportional to \eqref{App:eq.TV2}. Let us therefore consider the following Feynman diagram
  \begin{align}
  \raisebox{-0.3cm}{\includegraphics[width=2.5cm]{vsvv2}},
  \end{align}
  whose symmetry factor is $1$, but the corresponding multiplicity is actually $2$.
   This corresponds to the following two diagrams
  \begin{align}
  \raisebox{-0.95cm}{\includegraphics[width=3.2cm]{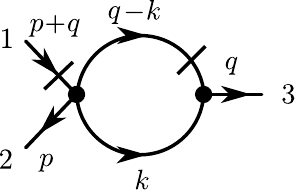}} \ =&\ [(d (d+4)+2) q_{3} \delta_{12} 
  \nonumber \\
  & + d (q(2) \delta_{13}+q_{1} \delta_{23})] C, \label{eq3} \\
  \raisebox{-0.95cm}{\includegraphics[width=3.2cm]{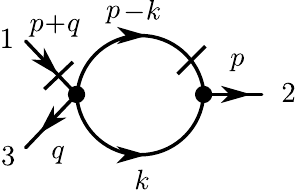}} \ =&\ [d ((d+4) p(2) \delta_{13} + p_{3} \delta_{12}
  \nonumber \\
  & + p_{1} \delta_{23} )+2 p_{2} \delta_{13}] C, \label{eq4}
\end{align}  
where $C$ stands for the following expression
\begin{equation}
  C = i\lambda \mu^{\varepsilon/2}\nu  \frac{(1-d)g_{1}}{12d(d+2)\varepsilon}
    \bigg( \frac{\mu^{2}}{\tau} \bigg)^{\varepsilon/2}.
\end{equation}
Contracting the sum of diagrams \eqref{eq3} and \eqref{eq4} with three transversal projection operators, we finally arrive at the result
  \begin{align}
  2 \ \raisebox{-0.5cm}{\includegraphics[width=2.5cm]{vsvv2}} = i\nu\lambda \mu^{\varepsilon/2} \XX_{123}  \frac{g_{1}}{24\varepsilon} \left( \frac{\mu^{2}}{\tau} \right)^{\varepsilon/2}.
  \end{align}


  \subsection{Four point vertex}
  
  The calculation of diagrams with four external lines proceeds in the same fashion as in the case of diagrams with three external lines. 
  All possible permutations of external points have to be considered, for instance the first diagram from  \eqref{1:eq.Gvsvvv} reads 
 
  \begin{align}
 3 \times \raisebox{-0.5cm}{\includegraphics[width=2.5cm]{vsvvv1}} & = 
 \raisebox{-0.55cm}{\includegraphics[width=3cm]{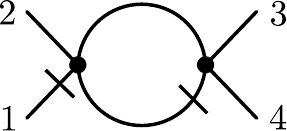}} 
 \nonumber\\ & +  \raisebox{-0.55cm}{\includegraphics[width=3cm]{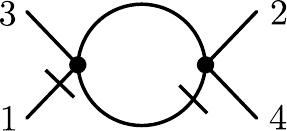}}
 \nonumber\\ & +  \raisebox{-0.55cm}{\includegraphics[width=3cm]{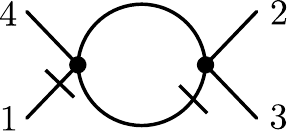}}. \label{app:diag4_perm}
 \end{align} 
 For illustration purposes we explicitly evaluate this diagram in the PRL limit. The general structure of the diagram is
 \begin{equation}
 \raisebox{-0.7cm}{\includegraphics[width=3cm]{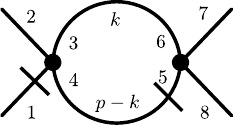}},
 \end{equation}
 where $ p $ is the sum of all external momenta flowing from the left to the right. The momentum space representation of the diagram is given by the expression
 \begin{align}
\int\displaylimits_{\kk,\omega} \frac{\nu_{0} \bar{g}_{10} \FF_{1234} \PP_{36}(\kk)\PP_{45}(\kk) \FF_{4567}}{[i\omega + \bar{\nu}_{0}(B_{\kk}+\bar{\tau}_{0})][\omega^{2} + \bar{\nu}_{0}^{2} (B_{\kk}+\bar{\tau}_{0})^{2}]},
 \end{align}
 where $ B_{\kk} = k^{2(1-\alpha)} $, and we have set the external momenta and frequency to zero because the diagram is already logarithmically divergent at $ d = d_{c} $. Performing the frequency integral, contracting indices and summing over permutations of external indices as in \eqref{app:diag4_perm}, we obtain
 \begin{align}
 \nu \bar{g}_{1} \FF_{1278} \frac{(d-1)(d^{2}+10d+12)}{12 d (2+d)} \int_{\kk} \frac{1}{(B_{\kk} + \tau)^{2}},
 \end{align}
 where we have used formulas \eqref{app:diag_momInt1} and \eqref{app:diag_momInt2}. In order to evaluate the above diagram, the following formula is useful
 \begin{align}
 \frac{1}{2S_{d}} \int \dd^{d}k \ f(|\kk|^{\sigma}) = \frac{1}{\sigma S_{2d/\sigma}} \int \dd^{2d/\sigma} k \ f(|\kk|^{2}). \label{app:FintA}
 \end{align}
 Using \eqref{app:Fint} with \eqref{app:FintA}, the result finally reads
 \begin{align}
 \nu \bar{g}_{1} \FF_{1278} \frac{(3 - 4 \alpha) \left(17 -18 \alpha + 4\alpha^{2}\right)}{24 (1-\alpha) (3 - 2 \alpha)} \left( \frac{\mu^{2 (1-\alpha)}}{\tau} \right)^{\frac{\tilde{\varepsilon}}{2 (1-\alpha)}},
 \end{align}
  where $ \tilde{\varepsilon} = 4(1-\alpha) - d $, we have set $ d = 4(1-\alpha) $ everywhere apart of the pole, and again rescaled the charge according to $ S_{d}\bar{g}_{1}/(2\pi)^{d} \rightarrow \bar{g}_{1} $.

	\section{Renormalization constants and anomalous dimensions \label{app:RC}}
	
	The explicit form of the renormalization constants for the model of active fluid with the LR interactions in the
	 minimal subtraction scheme is the following (the SR model corresponds to the case $ l = 0 $) 
		\begin{align}
	Z_{1} =&\ Z_{6} = 1, \\
	Z_{2} =&\ 1 - \frac{\bar{g}_{2}}{8} \sum_{l=0}^{\infty} \binom{-2}{l} \frac{\bar{w}^{l}}{\varepsilon+2\alpha l}, \\
	Z_{3} =&\ 1 +  \frac{3\bar{g}_{1}}{8} \sum_{l=0}^{\infty} \binom{-2}{l} \frac{\bar{w}^{l}}{\varepsilon+2\alpha l} ,\\
	Z_{4} =&\ 1 + \frac{5\bar{g}_{1}}{36}  \sum_{l=0}^{\infty} \binom{-2}{l} \frac{\bar{w}^{l}}{\varepsilon+2\alpha l} , \\
	Z_{5} =&\ 1 + \frac{17\bar{g}_{1}}{24} \sum_{l=0}^{\infty} \binom{-2}{l} \frac{\bar{w}^{l}}{\varepsilon+2\alpha l},
	\end{align}
	where $ g_{2} = \lambda^{2}, $ and we have performed a substitution $ g_{i} \bar{S}_{d} \rightarrow g_{i} $, with $ \bar{S}_{d} = S_{d}/(2\pi)^{d} $
	 and $ S_{d} $ being the surface of $ d $-dimensional sphere.
	
	The anomalous dimensions are obtained in a standard way from Eq.~\eqref{1:eq.gBeta}
	\begin{align}
	\gamma_{\bar{v}} &= - \gamma_{\bar{v}'} = \frac{\bar{g}_2}{16 (\bar{w}+1)^2}, \\
	\gamma_{\bar{g}_{1}} &= -\frac{17 \bar{g}_1 + 6 \bar{g}_2}{24 (\bar{w}+1)^2}, \\
	\gamma_{\bar{g}_{2}} &= -\frac{20 \bar{g}_1 + 27 \bar{g}_2}{72 (\bar{w}+1)^2}, \\
	\gamma_{\bar{w}} &= -\frac{\bar{g}_2}{8 (\bar{w}+1)^2}, \\
	\gamma_{\bar{\nu}} &= \frac{\bar{g}_2}{8 (\bar{w}+1)^2}, \\
	\gamma_{\bar{\tau}} &= -\frac{3 \bar{g}_1 + \bar{g}_2}{8 (\bar{w}+1)^2},
	\end{align}
	where the SR model corresponds to the case $ \bar{w} = 0 $.

	In the PLR limit, i.e. Eq. \eqref{eq:rescaled_func}, the only one loop contributions renormalize the mass term and the four-point interaction
	\begin{align}
	Z_{1} &= Z_{6} = 1, \\
	Z_{3} &= 1 + \frac{(3-2\alpha)(3-4\alpha)}{24(1-\alpha)} \ \frac{1}{\tilde{\varepsilon}}, \\
	Z_{5} &= 1 + \frac{(3 - 4 \alpha) \left(17 -18 \alpha + 4\alpha^{2}\right)}{24 (1-\alpha) (3 - 2 \alpha)} \ \frac{1}{\tilde{\varepsilon}}.
	\end{align}

  \bibliographystyle{apsrev}
  \bibliography{SBLH20} 

\end{document}